\documentstyle [12pt]{article}
\newcommand{\be}{\begin{equation}}
\newcommand{\ee}{\end{equation}}
\newcommand{\bea}{\begin{eqnarray}}
\newcommand{\eea}{\end{eqnarray}}
\newcommand{\ba}{\begin{array}}
\newcommand{\ea}{\end{array}}

\begin{document}
\baselineskip = 17 pt
 \title{
\vspace*{-2.5cm}
\begin{flushright}
\vspace{-0.3cm}
{\normalsize CERN-TH.7060/93} \\
\vspace{-0.3cm}
{\normalsize MPI-Ph/93-66} \\
\end{flushright}
\vspace{0.8cm}
Radiative Electroweak Symmetry Breaking and the Infrared
Fixed Point of the Top Quark Mass
 ~\\}
 \author{
 M. Carena$^{a,b}$, $\;\;\;$ M. Olechowski$^c$,\\
{}~\\
S. Pokorski$^a$\thanks{On leave of absence from the
Institute of Theoretical Physics, Warsaw University}
        $\;$ and  C. E. M. Wagner$^{a,b}$\\
 ~\\
$^a$Max-Planck-Institut
f\"{u}r Physik, Werner-Heisenberg-Institut\\
F\"{o}hringer Ring 6, D-80805 Munich, Germany.\\
{}~\\
$^b$ Theory Division, CERN, 1211 Geneva 23, Switzerland\\
{}~\\
$^c$Institute of Theoretical Physics, Warsaw University\\
ul. Hoza 69, 00-681 Warsaw, Poland\\
 }
\date{
\begin{abstract}
The infrared  quasi fixed point solution for the top quark mass
in the Minimal Supersymmetric Standard Model explains in  a natural
way large values of the top quark mass and appears as a prediction
in many interesting theoretical schemes. Moreover, as has been
recently pointed out, for moderate values of $\tan\beta$, in order
to achieve gauge and bottom-tau Yukawa coupling  unification,
the top quark mass  must be within $10 \%$ of its fixed point
value. In this work we show that the convergence of the top quark
mass to its fixed point value has relevant consequences for the
(assumed) universal soft supersymmetry breaking parameters at the
grand unification scale. In particular, we show that the low
energy parameters do not depend on $A_0$ and $B_0$ but on the
combination $\delta = B_0 - A_0/2$. Hence, there is a reduction
in the number of independent parameters. Most interesting,
the radiative $SU(2)_L \times U(1)_Y$ breaking condition implies
strong  correlations  between the supersymmetric mass
parameter $\mu$ and the
supersymmetry breaking parameters $\delta$ and  $M_{1/2}$ or $m_0$.
These correlations, which become stronger for $\tan\beta < 2$,
may have some fundamental origin, which would imply
the need  of a reformulation of the naive fine tuning criteria.\\
{}~\\
\begin{flushleft}
CERN-TH.7060/93\\
October 1993
\end{flushleft}
\end{abstract}}
\maketitle
\newpage
\baselineskip = 18 pt
\section{Introduction}

The increasing lower bound on the top quark mass has led
to a renewed interest in the  fixed point solutions
for the top quark Yukawa coupling \cite{IR}-\cite {Z}.
 In particular, the
infrared quasi fixed point  for the top quark mass
in the Minimal Supersymmetric Standard Model \cite{Dyn} appears
naturally in the framework of models with dynamical breaking
of the electroweak symmetry, the so-called top condensate
models. This is due to the fact that this solution
is associated with the renormalization
group trajectories on which the top quark Yukawa coupling, $h_t$,
becomes large, $Y_t \equiv h_t^2/4\pi = {\cal{O}}(1)$,
at scales of order $10^{16}$~GeV.
Most interesting, it has recently
been pointed out that, for moderate values of the ratio of
Higgs vacuum expectation values,  the condition of
bottom-tau
Yukawa coupling unification in minimal supersymmetric
grand unified theories \cite{Ramond}-\cite{DHR},
requires large values for the top quark Yukawa coupling at the
grand unification scale. This behaviour arises from the
necessity of contravening
the strong gauge coupling renormalization effects on the
bottom Yukawa coupling \cite{BABE}-\cite{LP}.
For the values of the
gauge couplings allowed  by  the most recent experiments
at LEP and from the grand unification condition, it follows
that the top quark
mass required to achieve bottom-tau Yukawa coupling
unification must be  within  10$\%$ of its infrared  quasi
fixed point value \cite{BCPW}.

The above predictions for the top quark mass
are independent of the source of the soft supersymmetry breaking
mass     parameters. In fact, since the only strong dependence
of the infrared quasi
fixed point prediction
on the supersymmetric spectrum
comes through the strong
gauge coupling $\alpha_3$,
this dependence can be characterized,
in grand unification scenarios,
by an effective
supersymmetric threshold scale $T_{SUSY}$ \cite{CPW}, \cite{LP},
which defines
the value of $\alpha_3(M_Z)$ for a fixed value of the
weak gauge couplings. The fixed point value for the top
quark mass  is given by
$M_t = C \sin\beta$, with $C \simeq 190$ - $210$ GeV for
$\alpha_3(M_Z) = 0.11$~-~$0.13$ and $\tan\beta$ being the ratio
of the vacuum expectation values of the Higgs fields.
For instance, a top quark mass
$M_t \leq 180$ GeV may only be obtained for $\tan\beta \leq 2$
(or for very large values of $\tan\beta$).

In the present work we shall analyse the potential implications
for the minimal supergravity model of the top quark mass being
at its infrared quasi fixed point value. In this model,
the low energy soft
supersymmetry breaking  mass parameters are thought to proceed
from common given values at the grand unification scale,
and the electroweak symmetry is broken radiatively.
As we shall show below, for low and
moderate values of
$\tan\beta$, the evolution of the soft
supersymmetry breaking parameters may be given as
a function of the ratio of the top quark Yukawa coupling
$Y_t$ to
its fixed point value $Y_f$,  yielding   definite analytical
predictions in the limit  $Y_t \rightarrow Y_f$.

In addition,  since
the low energy parameters must give
a proper breakdown of the $SU(2)_L \times U(1)_Y$ symmetry,
there is usually  some degree of fine tuning, which is
increased for low $\tan\beta \simeq 1$ as well as for very
large values
of $\tan\beta$. However, we shall show that, when the radiative
breaking condition on the supersymmetry breaking parameters is
imposed, it leads to relevant correlations
between the different high
energy parameters. These correlations
are stronger for exactly those values of  $\tan\beta$ for which
the naive fine tuning is strong. They may
have some fundamental explanation, which would make the usual
fine tuning argument inappropriate.  This result applies
in particular to the region $\tan\beta \leq 2$, which
corresponds to the infrared quasi fixed point values of the
top quark mass $M_t \leq 180$ GeV.

In the following, we shall perform
 a detailed analysis of the properties
mentioned above in the region of small and moderate values
of  $\tan\beta$. The large $\tan\beta$
region ($\tan\beta > 30$) will be analysed
in a forthcoming paper. In our present
study  we shall use the bottom - up
approach introduced in Ref. \cite{OP},
which enables a clear formulation
of these properties: while scanning  the whole low energy region in
our search for correlations of the mass parameters
at $M_{GUT}$ we can define the
exact patterns required to be close to the infrared quasi fixed point.
We shall  also make use of  analytical solutions for
 the low energy
parameters, which are extremely useful in understanding
the properties derived from the numerical
study.
In section 2 we give  analytical one loop expressions,
which show the dependence of the low energy scalar mass parameters
on the high energy soft SUSY
breaking mass parameters and on the top quark Yukawa coupling.
In section 3 we analyse the implications of the infrared fixed
point solution for the scalar mass parameter evolution. In section
4 we incorporate the radiative breakdown of the electroweak
symmetry and derive approximate analytical relations between the
soft supersymmetry breaking parameters at the high energy scale.
We compare our analytical  results with those we obtain from
the full numerical
computations, and we search
for  correlations between the different
high energy parameters. We reserve section 5 for our conclusions.

\section{Higgs Potential Parameters}

The  Higgs potential of the Minimal Supersymmetric
Standard Model may be written as \cite{Dyn}, \cite{CSW}-\cite{HH}
\begin{eqnarray}
V_{eff} & = & m_1^2 H_1^{\dagger} H_1 +
m_2^2 H_2^{\dagger} H_2 - m_3^2 (H_1^T i \tau_2 H_2
+ h.c.)
\nonumber\\
& + & \frac{\lambda_1}{2} \left(H_1^{\dagger} H_1 \right)^2
+ \frac{\lambda_2}{2} \left(H_2^{\dagger} H_2 \right)^2
+ \lambda_3 \left(H_1^{\dagger} H_1 \right)
 \left(H_2^{\dagger} H_2 \right)
+ \lambda_4 \left| H_2^{\dagger} i \tau_2 H_1^* \right|^2 ,
\end{eqnarray}
where at scales at which the theory is supersymmetric  the
running quartic couplings $\lambda_j$, with $j = 1 - 4$,
must satisfy the following conditions:
\begin{equation}
\lambda_1 = \lambda_2 = \frac{ g_1^2 + g_2^2}{4},\;\;\;\;\;
\lambda_3 = \frac{g_2^2 - g_1^2}{4},\;\;\;\;\;
\lambda_4 = - \frac{g_2^2}{2}.
\end{equation}
Hence, in  order to obtain the low energy values of the quartic
couplings, they must  be evolved using
the appropriate
renormalization group equations, as was explained in
Refs. \cite{CSW}-\cite{Chankowski}.
  The mass parameters $m_i^2$, with $i = 1$-$3$ must
also be evolved in a consistent way below the supersymmetry
breaking scale. The minimization conditions read
\begin{equation}
\sin(2\beta) = \frac{ 2  m_3^2  }{m_A^2}
\label{eq:s2b}
\end{equation}
\begin{equation}
\tan^2\beta = \frac{m_1^2 + \lambda_2 v^2 +
\left(\lambda_1
 - \lambda_2 \right) v_1^2}{m_2^2 + \lambda_2 v^2},
\label{eq:tb}
\end{equation}
where  $\tan\beta = v_2/v_1$, $v_i$ is  the vacuum expectation
value of the Higgs fields $H_i$, $v^2 = v_1^2 + v_2^2$,
$m_A$ is the CP-odd Higgs
mass,
\begin{equation}
m_A^2 = m_1^2 + m_2^2 + \lambda_1 v_1^2 +
\lambda_2 v_2^2 + \left( \lambda_3 + \lambda_4 \right) v^2
\end{equation}
and we define  the mass parameter $m_3^2$ to be positive.

The renormalization group equations for the mass parameters
may be found  in the literature \cite{Inoue}-\cite{Savoy2}.
Apart from the mass parameters $m_i^2$, appearing in the
effective potential, the evolution of the supersymmetric
mass parameter $\mu$ appearing in the superpotential $f$,
\begin{equation}
f =  h_t \epsilon_{ab} Q^b U H_2^a + \mu \epsilon_{ab}
H_1^a H_2^b ,
\label{eq:superp}
\end{equation}
is also relevant for the analysis of the radiative
electroweak symmetry breaking conditions.
In the above,
$Q^T = (T\;B)$ is the bottom-top left handed doublet superfield
and $U \equiv T^C$.
In Eq. (\ref{eq:superp})
we have just written the top quark Yukawa contribution, which is
the only one relevant for our analysis, since we are restricting
it to the region in which $\tan\beta$ takes small or moderate
values. The bilinear mass
term proportional to $m_3^2$ appearing in the Higgs
potential may
be rewritten as a soft supersymmetry breaking
parameter $B$ multiplied
by the Higgs bilinear term appearing in the
superpotential, that
is $m_3^2 = B \mu$. Analogously, the full
scalar potential  contains
scalar trilinear supersymmetry
breaking terms  with couplings $A_f$, proportional to
the terms in the superpotential associated with the
Yukawa couplings $h_f$.

The minimal supergravity model is obtained by assuming the
universality of the soft supersymmetry breaking parameters
at the grand unification scale: common soft supersymmetry
breaking mass terms $m_0$ and $M_{1/2}$ for the scalar and
gaugino sectors of the theory, respectively, and a common
value $A_0$ for all trilinear couplings $A_f$.
At the grand unification scale, the
mass parameters $B$ and $\mu$ take values $B_0$ and $\mu_0$,
respectively. Knowing the values of
the mass parameters at the unification scale, their low energy
values may be specified by their renormalization group evolution.
In the region of small and moderate values of $\tan\beta$, for which
the bottom and tau Yukawa coupling effects may be safely neglected,
an analytical solution for the evolution of the mass parameters may
be obtained,
for any given value of the top quark Yukawa coupling.

The solution
for the top quark Yukawa coupling, in terms of $Y_t$, reads
\cite{Ibanez},\cite{Savoy2}:
\begin{equation}
Y_t(t) = \frac{ 2 \pi Y_t(0) E(t)}{ 2 \pi + 3 Y_t(0) F(t)} ,
\end{equation}
with $E$ and $F$ being functions of the gauge couplings,
\begin{equation}
E = (1 + \beta_3 t)^{16/3b_3}
(1 + \beta_2 t)^{3/3b_2}
(1 + \beta_3 t)^{13/9b_1},
\;\;\;\;\;\;\;\;\;\;\;\; F= \int_{0}^t E(t') dt',
\end{equation}
 where $\beta_i = \alpha_i(0) b_i/4\pi$, $b_i$ is  the
beta function coefficient of the gauge coupling $\alpha_i$ and
$t = 2 \log(M_{GUT}/Q)$.
 As we mentioned above,
the fixed point solution is obtained for values of the
top quark Yukawa coupling that become large at the grand
unification scale, that is, approximately
\begin{equation}
Y_f(t) = \frac{2 \pi E(t)}{3 F(t)}.
\end{equation}
{}From here, considering the renormalization group
evolution of the mass parameters \cite{Ibanez}-\cite{BG},
the following
approximate analytical solutions  are obtained,
\begin{equation}
m_{H_1}^2 = m_0^2 + 0.5        M_{1/2}^2 \;\;\;\;\;   , \;\;\;\;\;
m_{H_2}^2 = m_{H_1}^2 + \Delta m^2\; ,
\label{eq:m12}
\end{equation}
where  $m_i^2 = \mu^2 + m_{H_i}^2$, with $i = 1,2$, and
\begin{eqnarray}
\Delta m^2 &  = & - \frac{3 m_0^2}{2} \frac{Y_t}{Y_f} + 2.3 A_0 M_{1/2}
\frac{Y_t}{Y_f} \left( 1 - \frac{Y_t}{Y_f} \right)
\nonumber\\
& - &
\frac{A_0^2}{2} \frac{Y_t}{Y_f} \left( 1 - \frac{Y_t}{Y_f} \right)
+ M_{1/2}^2
\left[
- 7 \frac{Y_t}{Y_f} + 3
\left(
\frac{Y_t}{Y_f} \right)^2 \right] \; .
\label{eq:dm}
\end{eqnarray}
Moreover, the  renormalization group
evolution for the supersymmetric mass parameter $\mu$
reads,
\begin{equation}
\mu^2 = 2 \mu_0^2 \left( 1 - \frac{Y_t}{Y_f} \right)^{1/2}  \; ,
\label{eq:mu}
\end{equation}
while the running of the soft supersymmetry breaking bilinear
coupling gives,
\begin{equation}
B = B_0 - \frac{A_0}{2} \frac{Y_t}{Y_f} + M_{1/2} \left(1.2
\frac{Y_t}{Y_f} - 0.6 \right).
\label{eq:b0}
\end{equation}
Finally, it is  also useful to present the evolution of the
supersymmetry breaking mass parameters of the supersymmetric partners
of the left and right handed top quarks,
\begin{equation}
m_Q^2 = 7.2 M_{1/2}^2 + m_0^2 + \frac{\Delta m^2}{3}  \;\;\;\;\; ,
\;\;\;\;\;
m_U^2 = 6.7 M_{1/2}^2 + m_0^2 + 2 \frac{\Delta m^2}{3},
\label{eq:sqm}
\end{equation}
respectively. We shall concentrate on the renormalization
group evolution of the   supersymmetric mass parameters
given above, Eqs. (\ref{eq:m12}) - (\ref{eq:sqm}), since
within the bottom - up approach introduced
in Ref. \cite{OP}
these are sufficient for the determination of the high
energy parameters.

 A remark is in order. The above solutions have been obtained
by using the perturbative
one loop renormalization group equations for
the gauge and Yukawa couplings, as well as
for the mass parameters.
Hence, they can only be used for values of
the top quark Yukawa coupling at
the grand unification scale  within the range of validity
of perturbation theory, $Y_t(0) \leq 1$. Since there is a
one to one relationship between the values of $Y_t(0)$ and
the degree of convergence of the top quark Yukawa coupling
to its infrared quasi fixed point value, this bound
implies that the solutions associated with top quark
Yukawa couplings that are closer than $0.5 \%$ to the
fixed point value cannot be studied within the one loop
approximation. In the following, when talking about the limit
$Y_t \rightarrow Y_f$, we will be implicitly assuming that we
are working within the range of validity of perturbation theory.
In addition, for values of $Y_t$ that are very
close to its fixed point value, two loop effects may
become important. Therefore, in our numerical solution we have
considered the full two loop renormalization group evolution
for gauge and Yukawa couplings.

 The coefficients characterizing the
 dependence of the mass parameters on the universal gaugino
mass $M_{1/2}$ depend
on the exact value of the gauge
couplings. In the
above, we have taken the values of the coefficients that
are  obtained for $\alpha_3(M_Z) \simeq 0.12$.
The above  analytical solutions are    sufficiently accurate for
the purpose of        understanding    the properties of
the mass parameters in the limit $Y_t \rightarrow Y_f$. We shall then
confront the results of our analytical study with
those obtained from the numerical two loop analysis.

\section{Properties of the Fixed Point Solutions}

The above expressions
 show important properties of the solution
when $Y_t \rightarrow Y_f$:\\
{}~\\
a) The mass parameters  $m_{H_2}^2$, $m_Q^2$ and
$m_U^2$ become  very weakly dependent on the supersymmetry
breaking parameter $A_0$. In fact, the dependence on $A_0$ vanishes
in the formal limit $Y_t \rightarrow Y_f$.
The only relevant dependence on $A_0$
enters through the mass parameter $m_3^2$. This leads to
property (b). \\
{}~\\
b) There is an effective reduction in
the number of free independent soft
supersymmetry breaking parameters. In fact, the dependence on
$B_0$ and $A_0$ of the low energy solutions is effectively
replaced by
a dependence on the parameter
\begin{equation}
\delta = B_0 - \frac{A_0}{2}.
\end{equation}
c)  From Eq. ( \ref{eq:mu}), it follows that
the coefficient relating $\mu$ to $\mu_0$ tends  to
zero as $Y_t  \rightarrow Y_f$. This tendency is, however, much
slower than that  of the coefficient associated
with the $A_0$ dependence of the mass parameters leading to
properties (a) and (b). For instance, even if $Y_t$ lies
as close as only $0.5 \%$ away from $Y_f$, this coefficient is
still of order one.  The relevant property following from
Eq. (\ref{eq:mu}) is that, for the same low energy value of
$\mu$, consistent with the radiative breaking of $SU(2)_L \times
U(1)_Y$, $\mu_0$ should scale like $\left(1 - \frac{Y_t}{Y_f}
\right)^{-1/4}$.\\
{}~\\
d) There is a  very interesting dependence of the low energy
mass parameters on $m_0$. For example, the $m_0$ dependence
of the combination $m_Q^2 +m_{H_2}^2$ vanishes in the formal
limit $Y_t \rightarrow Y_f$. Moreover,
the right stop mass $m_U^2$ becomes itself independent of
$m_0^2$ in this limit.\\

One remark is in order. As we said above, the explicit dependence
on $A_0$ vanishes as  the top quark Yukawa coupling approaches
its infrared quasi fixed point, and it is
replaced by a dependence on the parameter $\delta$. However,
as we discussed above, we can only make a reliable perturbative
analysis of solutions  for
which the top quark Yukawa coupling is very close to,
but not exactly at, its infrared quasi fixed point value.
For these
solutions, the explicit dependence on $A_0$ of the mass parameters
is negligible,
 within a certain range of values for $A_0$, which
depends on how close to one is the ratio $Y_t/Y_f$. For instance,
if $Y_t$ is at most about ten (one) per cent away from its
quasi
fixed point value, the dependence on $A_0$ is negligible for
$A_0^2$ taking values smaller than one
(two) order(s) of magnitude of the value of
the soft supersymmetry breaking parameters $m_0^2$ and
$M_{1/2}^2$. For still larger values of $A_0$ the explicit
dependence on this parameter may, in principle, reappear.
Very large values of $A_0$
are, however, restricted by the  condition ensuring
the absence of
a colour breaking minimum at $M_{GUT}$ \cite{ILEK},
\begin{equation}
A_0^2 \leq 3 (3 m_0^2 + \mu_0^2).
\label{eq:color}
\end{equation}
It is clear that values of $A_0$  much larger than $m_0$
and $M_{1/2}$  are consistent with Eq. (\ref{eq:color})
only for $\mu$ (which is of order $\mu_0$) much larger
than the universal scalar and gaugino masses.
Actually, as will be discussed below,
the condition $\mu \gg m_0,M_{1/2}$ is
consistent with the
requirement of radiative breaking of the electroweak symmetry
for $\tan\beta$ close to one.
In the following, we will
always assume that for values of $\tan\beta$ close to one,
we are sufficiently close to the fixed point  so that the
explicit dependence of the mass parameter $m_{H_2}^2$
on $A_0$ may be neglected for any $A_0$ consistent with
Eq. (\ref{eq:color}). This is naturally the case
in the numerical solutions we studied. Moreover, we would
like to stress that even for $\tan\beta$ closer to one
and/or values
of $Y_t$ further away from its fixed point value,
there is always
an interesting range of values for $A_0$ where the explicit
dependence on this parameter can be neglected (although this
dependence may
reappear by chosing $A_0^2$ close to its
upper bound, Eq. (\ref{eq:color})).

\section{Radiative Breaking of $SU(2)_L \times U(1)_Y$}

In general,
the soft supersymmetry breaking parameter space is
subject to experimental and theoretical constraints. The
experimental constraints come from the present lower
bounds on the supersymmetric particle masses. For example
the present lower bound on the gluino mass  implies a
lower bound on the soft supersymmetry breaking parameter
$M_{1/2}$
\footnote{ The above holds only in the case when
one ignores the
possibility of a light gluino window. However, the light gluino scenario
is  ruled out when asking
for radiative breaking  of $SU(2)_L \times U(1)_Y$ for
the infrared fixed point solution, unless one is willing to relax
the condition of universality of the soft supersymmetry breaking scalar
masses at the unification scale \cite{Nanop}, \cite{CCMNW}.}.
On the other hand, one should require the
stability of the effective potential and  a
proper breaking of the $SU(2)_L \times U(1)_Y$ symmetry.
There are other theoretical constraints, for example those
coming from the degree of fine tuning of a given solution.
Although a rigorous definition of this concept is lacking,
different numerical ways of measuring the degree of fine
tuning have been proposed in the literature. In general,
independent parameters are assumed. If there were, however,
some interrelation between different parameters coming from
the fundamental dynamics leading to the soft supersymmetry
breaking terms, it would show up in the form of strong correlations
between these parameters in the radiative breaking solutions.
Hence, if strong correlations are found, the naive fine tuning
criteria may be inappropriate for the analysis of the degree of
fine tuning of a given solution.

In this work, we have performed
a complete numerical analysis
of the constraints coming from the requirement of a proper
radiative breaking of $SU(2)_L \times U(1)_Y$, the results of
which are shown in Figs. 1 to 5.
In order to get an analytical
understanding of the properties derived numerically, it is
most useful to present
an approximate theoretical analysis of the radiative
breaking condition. From the  expressions for
the mass parameters obtained above in
Eqs. (\ref{eq:m12})-(\ref{eq:sqm}), together with the
minimization condition in Eq. (\ref{eq:tb}), and ignoring
at this level the radiative corrections  to
the quartic couplings, it follows that
\begin{equation}
\mu^2 + \frac{M_Z^2}{2} =
m_0^2 \frac{
1 + 0.5 \tan^2 \beta }{ \tan^2 \beta -1} +
M_{1/2}^2
\frac{ 0.5 + 3.5 \tan^2 \beta}
{\tan^2 \beta -1} .
\end{equation}
Thus, as we mentioned in the previous section,
in the limit $\tan\beta \rightarrow 1$, we find $\mu^2 \gg
m_0^2, M_{1/2}^2$. In addition,
the above  implies that, for a fixed value of $\tan\beta$ and
$M_{1/2} > M_Z,m_0$, there is a strong
correlation between $\mu$ and $M_{1/2}$, which is approximately
given by
\begin{equation}
\mu^2 \simeq  M_{1/2}^2
\frac{ 0.5 + 3.5 \tan^2\beta}
{\tan^2\beta -1}.
\label{eq:mug}
\end{equation}
If, instead,  $m_0^2 \gg M_{1/2}^2,M_Z^2$, a linear
correlation between $m_0$ and $\mu$ is obtained,
\begin{equation}
\mu^2 \simeq
m_0^2 \frac{
1 + 0.5 \tan^2 \beta
}{ \tan^2 \beta -1} .
\end{equation}
The correlations are
stronger, for lower values of $\tan\beta$, and become
almost exact for $\tan\beta \rightarrow 1$.
The inclusion of the radiative
corrections to the Higgs quartic couplings
give corrections to the effective mass parameters squared,
which are of order $M_Z^2$ and hence
do not modify the above behaviour.

The numerical results for the correlations ($m_0 - \mu$) and
($M_{1/2} - \mu$), which follow from the requirement of a proper
radiative electroweak breaking, are shown in Figs. 1
and 2. We choose  values of $M_t$ and $\tan\beta$ for which
the top quark Yukawa coupling is one to three per cent away from
its infrared quasi fixed point value.
In order to fully understand those plots we note that, in the
limit $Y_t \rightarrow Y_f$, the following relations hold:
\begin{equation}
m_U^2 \simeq 4 M_{1/2}^2 ,\;\;\;\;\;\;\;\;\;
m_0^2 \simeq 2 m_Q^2 - 3 m_U^2.
\end{equation}
Thus, as long as we are interested in
$SU(2)_L \times U(1)_Y$ breaking, with the common
soft supersymmetry
breaking parameters
such as to give low energy squark masses  that are
below some common upper bound (which is taken to be
1 TeV in Figs. 1 and 2), then
the upper limits for the soft supersymmetry breaking
gaugino and
scalar masses,
$M_{1/2}^U$ and $m_0^U$, respectively, satisfy the relation
$m_0^U \simeq 2 \sqrt{3} M_{1/2}^U$.
Therefore, in a large portion
of the allowed parameter space $m_0^2 \gg M_{1/2}^2$ and,
as seen in Fig.1, the correlation ($m_0 - \mu$) is the dominant,
gross feature of the obtained solutions. A closer look at
Fig. 1 shows, however, that this correlation is sharper for larger
values of $m_0$ (in particular for low values of $\tan\beta$)
and gradually disappears for small $m_0$ (i.e. in the region where
we expect the ($M_{1.2} - \mu$) correlation predicted for the
regime $M_{1/2} > m_0$). The correlation between $\mu$ and
$M_{1/2}$ is, instead, not explicit in Fig. 2.
Again, this is just a reflection of the fact that most of
the solutions in Figs. 1 and 2 satisfy the relation
$m_0 > M_{1/2}$.

The correlation ($M_{1/2} - \mu$) becomes
sharply visible in the subset of solutions with $M_{1/2} >
m_0$. It is very interesting that for $\tan\beta \leq 2$ such
solutions (and only those) are selected by the physical
requirement of the acceptable neutralino relic abundance,
$\Omega h^2 \leq 1$. It is well known that in the
minimal supergravity model the neutralino relic abundance
is precisely calculable, with no additional free parameters.
In Fig. 3 we show the
solutions to the electroweak radiative breaking, for
$\tan\beta = 1.2$ and $M_t = 160$ GeV, which satisfy the
constraint $0.1 \leq \Omega \leq 0.7$. The calculation is
based on the formulation of Ref. \cite{Gondolo}.
First, we compare the available range in $M_{1/2}$ and
$m_0$ without and with the cut on $\Omega h^2$ and observe
that the $\Omega$ cut gives a strong upper limit on $m_0$ and selects
solutions with $M_{1/2} \geq m_0$. This upper bound on
$m_0$ has a very simple qualitative explanation. The low
$\tan\beta$ solutions to radiative breaking always give
$\mu > M_{1/2}$ (as follows from
Eq. (\ref{eq:mug}) and is seen in Fig. 2)
and, consequently, the predicted lightest
neutralinos are strongly
bino-like. Their annihilation then proceeds mainly
via slepton exchange and the requirement of the acceptable
relic abundance (which is inversely proportional to the
annihilation cross section) puts an upper bound on the
slepton mass, i.e., also on the grand unification
parameter $m_0$.
Then, the other features seen in Fig. 3  follow
naturally: strong
($M_{1/2} - \mu$) correlation and no $(m_0 - \mu)$
correlation.

The second radiative breaking condition, Eq. (\ref{eq:s2b}),
leads to the following
relation
\begin{equation}
\sin 2 \beta \left( 2 \mu^2 + \frac{m_0^2}{2} - 3 M_{1/2}^2
\right) = 2  \mu \left(\delta + 0.6 M_{1/2}\right).
\label{eq:sin2}
\end{equation}
Additional properties of the solution may be obtained by using
Eq. (\ref{eq:sin2}).
As we mentioned above, we define the mass parameters $m_3^2$,
$m_0$ and $M_{1/2}$ to
be positive. With this sign convention, there are two
different regimes, depending on the sign of the supersymmetric
mass parameter $\mu$.
Since, for example,
in the region $M_{1/2} > M_Z,m_0$ there is a strong
correlation, Eq.(\ref{eq:mug}),
between $\mu$ and $M_{1/2}$,
Eq. (\ref{eq:sin2})
shows that for $\mu > 0$
there is  a strong linear correlation
between the parameters $\delta$ and $M_{1/2}$,
\begin{equation}
\delta \simeq  M_{1/2}
\left( - 0.6 + \frac{ 2 \sin 2 \beta \left( 1 + \tan^2 \beta
\right) }{\sqrt{
\left( 0.5 + 3.5 \tan^2 \beta \right) \left(
\tan^2 \beta -1 \right) }} \right),
\label{eq:deg1}
\end{equation}
and also between    $\mu$ and $\delta$,
\begin{equation}
\delta = \mu \frac{ \left( - 0.6 \sqrt{
\left( 0.5 + 3.5 \tan^2 \beta \right) \left(
\tan^2 \beta -1 \right) } + 2 \sin(2\beta)
\left( 1 + \tan^2 \beta \right) \right) }
{ 0.5 + 3.5 \tan^2 \beta} .
\label{eq:deltam0}
\end{equation}
Analogously, in the region $m_0 \gg M_{1/2}$, where the linear
correlation between $m_0$ and $\mu$ holds, we obtain
a strong linear correlation between $\delta$ and $m_0$
\begin{equation}
\delta \simeq m_0 \frac{ \sin(2\beta) 0.75 (1 + \tan^2 \beta) }
{ \sqrt{ \left(1 + 0.5 \tan^2 \beta \right)
\left( \tan^2 \beta -  1 \right) }},
\label{eq:dem01}
\end{equation}
as well as a different
correlation between $\mu$ and $\delta$,
\begin{equation}
\delta = \mu \frac{ \sin(2 \beta) 0.75 \left( 1 +
\tan^2 \beta \right)}{ 1 + 0.5 \tan^2 \beta}.
\label{eq:demum0}
\end{equation}

For $\mu \leq 0$, instead, the correlation between $\delta$
and $M_{1/2}$ in the $M_{1/2} > m_0,M_Z$ regime reads
\begin{equation}
\delta \simeq  - M_{1/2}
\left(  0.6 + \frac{ 2 \sin 2 \beta \left( 1 + \tan^2 \beta
\right) }{\sqrt{
\left( 0.5 + 3.5 \tan^2 \beta \right) \left(
\tan^2 \beta -1 \right) }} \right).
\label{eq:deg2}
\end{equation}
It is interesting to compare Eqs. (\ref{eq:deg1}) and
(\ref{eq:deg2}). A variation in the sign of $\mu$ yields a
different absolute value of the coefficients relating
$\delta$ to $M_{1/2}$. Hence, the resulting correlations are
not symmetric under a change in sign of $\delta$.
The linear correlation between $\delta$ and $\mu$ in this
regime is hence given by
\begin{equation}
\delta  \simeq \mu \frac{ \left(  0.6 \sqrt{
\left( 0.5 + 3.5 \tan^2 \beta \right) \left(
\tan^2 \beta -1 \right) } + 2 \sin(2\beta)
\left( 1 + \tan^2 \beta \right) \right) }
{ 0.5 + 3.5 \tan^2 \beta}.
\label{eq:delmu}
\end{equation}
Furthermore, in
the region $m_0^2 \gg M_{1/2}^2$, the following
linear correlation
between $\delta$ and $m_0$ is present for $\mu \leq 0$,
\begin{equation}
\delta \simeq -  m_0 \frac{ \sin(2\beta) 0.75 (1 + \tan^2 \beta) }
{ \sqrt{ \left(1 + 0.5 \tan^2 \beta \right)
\left( \tan^2 \beta -  1 \right) }}.
\label{eq:deltam00}
\end{equation}
The above expression
differs only in sign from the one obtained in
Eq. (\ref{eq:dem01}), for the regime $\mu \geq 0$, while
the resulting correlation between $\mu$ and $\delta$ is exactly
that obtained in Eq. (\ref{eq:demum0}).

Interestingly enough, although they have quite a different
dependence on $\tan\beta$, for $\mu \leq 0$,
the numerical values of
the coefficients relating $\delta$ with
$\mu$ in the two different regimes studied above,
Eqs. (\ref{eq:demum0}) and (\ref{eq:delmu}),  are remarkably
close to each other for $\tan\beta \leq 10$. Hence, for
those values of $\tan\beta$ a strong correlation between
$\delta$ and $\mu$ is  expected to appear, for $\mu \leq 0$,
for the whole range of values of $m_0$ and $M_{1/2}$.
Due to the numerical value of the coefficients, the
correlation between $\mu$ and $\delta$ in the regime
$\mu \leq 0$ should  improve for $\tan\beta $
close to 1 as well as for $\tan\beta$ close to 5.
This is actually observed in Fig. 4.

In the region $\mu \geq 0$, instead, the numerical values of
the two coefficients relating $\mu$ and $\delta$
only coincide in the limit $\tan\beta
\rightarrow 1$, becoming quite different as $\tan\beta$ increases.
Hence, for $\mu \geq 0$
we expect the correlation ($\mu - \delta$)
to be good only for small values of
$\tan\beta \simeq 1$.  On the contrary,
for larger values of $\tan\beta$
this linear  correlation
is lost. This is due
to the fact that the coefficients in both
regimes are quite different and, in addition, the correlation
of $\mu$ with $M_{1/2}$ and $m_0$ in the two studied regimes
becomes weaker than for lower values of $\tan\beta$.  For
instance, for $\tan\beta = 10$ the correlations observed for
smaller values of $\tan\beta$ almost disappear,  as  is
shown in Fig. 4.

A very similar analysis applies to the correlation between
the pseudoscalar mass $m_A$ and the parameter $\delta$.
Combining Eqs. (\ref{eq:m12}) and (\ref{eq:dm}) with
Eqs. (\ref{eq:deg1}) and (\ref{eq:deltam0})
(or (\ref{eq:deg2}) and (\ref{eq:deltam00})) it is easy
to find that, in the regime $M_{1/2} > m_0$, the linear
relation
\begin{equation}
m_A = - \delta
\frac{ \sqrt{ 2 ( 1 + 7 \tan^2\beta )
(1 + \tan^2\beta) }}{
0.6 \sqrt{ (0.5 + 3.5 \tan^2\beta) (\tan^2\beta - 1) }
+ (-)
4 \tan\beta }
\end{equation}
appears for $\mu < 0$ ($\mu > 0$), while for
$m_0^2 \gg M_{1/2}^2$ one finds
\begin{equation}
m_A = - (+)  \delta  \sqrt{
\frac{ 3 (1 + \tan^2\beta ) ( 2 + \tan^2\beta)}
{ 9 \tan^2\beta}}.
\end{equation}
The properties of the correlation between $m_A$ and $\delta$
are similar to those
 ones of the ($\delta - \mu$) correlation.
Indeed,
in the limit $\tan\beta \rightarrow 1$ there
exists a strong linear correlation between $m_A$ and
$\delta$, in both regimes, $M_{1/2} > M_Z,m_0$ and $m_0
\gg M_{1/2}$. This is clearly seen in the numerical
results shown in Fig. 5. As before, this correlation
is stable for small and moderate values of $\tan\beta$
in the $\mu < 0$ regime, while it rapidly disappears
in the $\mu > 0$ regime when $\tan\beta$ becomes
larger than 1. In general, the correlation
becomes weaker for increasing values of $\tan\beta$.

One  remark should be made regarding the condition
$\delta = 0$. Since at the infared quasi fixed point
the explicit dependence on the $A_0$ parameter is
replaced by the dependence on $\delta$, finding
solutions that satisfy the relation $\delta = 0$ would
also imply to have solutions for $A_0 = B_0 = 0$.
The condition $\delta = 0$ cannot be fulfilled for
values of $\mu \leq 0$, for $\tan\beta < 10$,
since the strong correlation between
$\mu$ and $\delta$ renders this impossible. For
$\mu > 0$ and moderate values of $\tan\beta$,
where the correlation between $\delta$ and $\mu$ is
lost, this condition may be achieved. Indeed, the complete
study of the minimization condition shows that $\delta = 0$
is achievable for $\tan\beta \geq 4$ for positive values of
the mass parameter $\mu$.

\section{Conclusions}

In this article, we have studied the properties of
the minimal supergravity model for the case in which
the top quark mass is close to its infrared quasi
fixed point solution. To study the regime of
the infrared quasi fixed point solution for $m_t$ is
of interest for various reasons. This solution explains
in a natural
way the relatively large value of the top quark mass.
Moreover, it appears as a prediction
in many interesting theoretical scenarios. In particular,
it has been shown that for the  presently allowed values of
the bottom quark mass and the electroweak gauge couplings,
the conditions of gauge and bottom-tau Yukawa coupling
unification imply a strong convergence of the top quark
mass to its infrared fixed point value.

Our main conclusion is that a proximity of the top
quark mass to its infrared quasi fixed point values would
have important implications for the mechanism of
radiative electroweak symmetry breaking in the minimal
supergravity model. In particular, we show that
there is a reduction in the number of independent parameters:
the low energy parameters do not depend on the
grand unification scale
parameters $A_0$ and $B_0$, but on the combination
$\delta = B_0 - A_0/2$. Furthermore, we prove the existence of
important correlations between the remaining free parameters of the
model, which emerge as the exact pattern of all solutions with
proper electroweak symmetry breaking. These correlations become
particularly strong in the low $\tan\beta$ region which
corresponds to the infrared fixed point values of the top
quark mass  $M_t \leq 180$ GeV. It is tempting to speculate
that they have some fundamental explanation. The usual criteria
of fine tuning, with all parameters treated as independent,
would then have to be abandoned.
{}~\\
\newpage
{}~\\
{\bf{Acknowledgements}}\\
{}~\\

M.O. and S.P. are grateful to P.H. Chankowski for numerous
discussions on radiative electroweak symmetry breaking and
to Z. Lalak for a useful comment. M.O. and S.P. are partially
supported by  the Polish Committee for Scientific Research.

\newpage
{}~\\
{\bf{FIGURE CAPTIONS}}
{}~\\
{}~\\
{\bf{Fig. 1.}} Solutions for the parameters $m_0$ and $\mu$
obtained by scanning the parameter space with the
requirement of a proper radiative electroweak symmetry
breaking solution, for four
different  values of $M_t$ and $\tan\beta$
consistent with the infrared quasi fixed point solution.
The scanning of solutions was performed considering
values of $m_A$, $m_Q$ and $m_U$ up to 1 TeV.
\\
{}~\\
{\bf{Fig. 2.}} Same as in   Fig. 1,
but for the radiative electroweak
symmetry breaking solutions in the
$M_{1/2} - \mu$ parameter space.\\
{}~\\
{\bf{Fig. 3.}} Analysis of the $\Omega$ cut effects in the
radiative electroweak symmetry breaking  solutions
in the mass parameter space, for a given set of values of $M_t$ and
$\tan\beta$ consistent with the infrared quasi fixed point
solution. \\
{}~\\
{\bf{Fig. 4.}} Same as in Fig. 1, but for the radiative electroweak
symmetry breaking solutions in the
$\mu - \delta$ parameter space.
\\
{}~\\
{\bf{Fig. 5.}} Same as in Fig. 1, but for the radiative electroweak
symmetry breaking
solutions in the
$m_A - \delta$ parameter space. \\
{}~\\
\end{document}